\documentclass[preprint,aps,showpacs]{revtex4}
\usepackage{amssymb}
\usepackage{amsmath}
\usepackage{epsfig}
\usepackage{bm}
\usepackage{latexsym}

\begin{document}

\title{Network of tRNA Gene Sequences }
\author{Fangping Wei\footnote{cocoapple@sjtu.edu.cn}, Sheng Li\footnote{lisheng@sjtu.edu.cn}
, Hongru Ma\footnote{hrma@sjtu.edu.cn}
 }

\date{\today}

\begin{abstract}
We showed in this paper that similarity network can be used as an powerful
tools to study the relationship of tRNA genes. We constructed a network of
3719 tRNA gene sequences using simplest alignment and studied its topology,
degree distribution and clustering coefficient. It is found that the
behavior of the network shift from fluctuated distribution to scale-free
distribution when the similarity degree of the tRNA gene sequences increase.
tRNA gene sequences with the same anticodon identity are more self-organized
than the tRNA gene sequences with different anticodon identities and form
local clusters in the network. An interesting finding in our studied is some
vertices of the local cluster have a high connection with other local
clusters, the probable reason is given. Moreover, a network constructed by
the same number of random tRNA sequences is used to make comparisons. The
relationships between properties of the tRNA similarity network and the
characters of tRNA evolutionary history are discussed.
\end{abstract}

\affiliation{Institute of Theoretical Physics, Shanghai Jiao Tong University, Shanghai
200240, China}
\maketitle



\section{Introduction}

Transfer ribonucleic acid, or tRNA for short, is an important molecule which
transmits genetic information from DNA to protein in molecular biology. It
has been known that all tRNAs share a common primary, secondary, tertiary
structure. Most tRNA sequences have a "CCA" hat in terminus 5' and a polyA
tail in terminus 3' in its primary structure. Its secondary structure is
represented by a cloverleaf. They have four base-paired stems and a variable
stem, defining three stem loops (the D loop, anticodon loop, and T loop) and
the acceptor stem, to which oligonucleotides are added in the charging step{%
\cite{Robert2001}}. Variable loop varies in length from 4 to 13 nt, some of
the longer variable loops contain base-paired stems. The tRNAs also share a
common three-dimensional shape, which resembles an inverted "L". Though much
effort had been put on tRNA research in the past time, little is known about
specific features of tRNA that are exclusive to a species, taxa or
phylogenetic domain level{\cite{Robert1966}}. With the progress of genome
projects, a vast amount of nucleotide sequence data of tRNA is now
available, which makes it possible to study the tRNA genes expression for a
wide range of organisms.

Recently scientists are trying to find specific feature in genes families by
a new tool---complex networks. With the development of techniques on
oligonucleotide or cDNA arrays, using gene chips to erect a complicated
network and studying its feature and evolution has become a hot subject, and
has gained a success {\cite{Himanshu2002, Stefan2001, Stefan2003, Jeong2001}}%
. Basically, the networks can be classified into two types in terms of its
degree distributions $p(k)$ of nodes: exponential networks and scale-free
networks. The former type has a prominent character that although not all
nodes in that kind of network would be connected to the same degree, most
would have a number of connections hovering around a small, average value,
i.e. $k\thicksim \langle k\rangle $, where $k$ is the number of edges
connected to a node and is called degree of the node. The distribution leads
to a Poisson or exponential distribution, such as random graph model{\cite%
{ERD 1960} }and small-world model{\cite{Watts1998}}, which is also called
homogenous networks. The latter type network has a feature that some nodes
act as "very connected" hubs which have very large numbers of connections,
but most of the nodes have small numbers of connections. Its degree
distribution is a power-law distribution, $p(k)\thicksim k^{-\gamma }$. It
is called inhomogeneous network, or scale-free network{\cite{Barabasi1999}}.

The tRNA sequences have similarities in sequences and structure, which make
it possible to construct networks and use specialized clustering techniques
to make classification. The similarity of tRNA sequences suggests their
relationships in evolutionary history. If we consider all the tRNA sequences
at present evolve from common ancestor via mutation, the sequence similarity
will reveal their evolutionary affiliation. There are lots of tRNA
sequences. The similarities of every two of the sequences are different.
Lots of data will be dealt with. Since complex network is a good model to
describe and study complex relationships, the network model may be useful in
this field. In this paper, we constructed a similarity network of 3917 tRNA
genes in order to show network model is a powerful tool to study the
evolutionary relationships among the tRNA genes. The topology of the network
is discussed, the degree distribution and clustering coefficient are
considered,and the network constructed by the same number of random tRNA
sequences is used to make comparisons.

\section{Materials and methods}

\subsection{tRNA sequences}

Transfer RNA sequence have been collected into database by Sprinzl et al{%
\cite{Sprinz1998}}in 1974. All of our data, 3719 tRNA genes sequences, are
retrieved from this database (free available at
http://www.uni-bayreuth.de/departaments/biochemie/sprinzl/trna/), which
including 61 anticodon subsets, 429 species, and 3 kingdoms: Archaea,
Bacteria, and Eucarya. Each tRNA sequence has 99 bases when the variable
stem is considered. For convenience of alignment, the absent bases in some
positions of the tRNAs are inserted with \textquotedblleft
blank\textquotedblright . Firstly. we align the tRNA sequences with the same
anticodons, and then align all 3719 tRNAs. Since there have been too many
conclusions proving that tRNA genes have a high similarity in sequences{\cite%
{Stefan1999, Sergey1993,Saks1998}}, the results of the alignment of 3719
tRNA gene sequences will not be listed in detail. We only focus on some
prominent characters of the statistics of the alignment.

\subsection{tRNA sequences network}

If each base including the inserted \textquotedblleft
blank\textquotedblright\ is considered equally, the length of a tRNA is $L=99
$. To align two tRNA sequences, a parameter $s$ is used to depict their
similarity degree, which indicates how many bases in the same position of
two tRNA gene sequences are identical. For example, if the first bases of
two tRNA sequences both are A, one score is added to $s$. Obviously, $0\leq
s\leq 99$. Although it is the simplest kind of alignment, as we show later,
it gives lots of information of the relationships among tRNA genes. When $%
s=99$, it means two sequences are matched perfectly. Since the perfectly
matched sequences have the the same significance in biology, we take only
one of them as a representative. To construct the tRNA similarity network,
every sequence is considered as a node. If the alignment score $s$ of two
tRNA sequences is larger than a given similarity degree $s_{0}$, put an edge
between the corresponding nodes. Obviously, if $s_{0}$ is small, the nodes
will connect closely, and when $s_{0}$ grows larger, the number of
connections will decrease.

For comparison, we make a similarity network of the same number of random
tRNA genes. To generate the random tRNA genes, every base of the sequences
is randomly taken from the four bases (C, G, A and T) and the sequences must
conform to the prototype of the real tRNA, which means the sequences we
generate randomly must confirm the secondary structure of tRNA.

\section{Graph tool}

Pajek (the Slovene word for spider), a program for large-network analysis{\
\cite{Batagel1998} }(free available at
http://vlado.fmf.uni-lj.si/pub/networks/pajek/), was used to map the
topology of the network.

\section{Results}

\subsection{The topology of network}

Figure \ref{figure1} displays several typical topologies of the similarity
network of different kinds of tRNA gene sequences. Figure \ref{figure1} (a),
(b) and (c) are similarity networks constructed by tRNA genes with the same
anticodons (CGC, CCA and TGC respectively) and $S_{0}=60$. The networks of
tRNA genes with the same anticodon identity are highly clustered. Some of
them divide into two or more clusters, such as figure \ref{figure1}(c). Each
of the clusters almost entirely connected when $s_{0}$ is small. When $s_{0}$
grows large, the connection number decreases, and the network becomes not so
closely connected. Figure\ref{figure1}(d) is the similarity network of
anticodon GTT when $s_{0}=80$. As more nodes added in the network, the
network becomes more complex. Figure \ref{figure1} (e), (f) shows the
network with a large $N$ (the number of nodes). (e) is the network
containing anticodons CAT and GCC with $S_{0}=80$, and (f) is the network of
all $3420$ tRNA sequences with $S_{0}=90$. Small local clusters with the
same anticodons get together to form a large cluster, "very connected" hubs\
can be observed in the center of the network (figure \ref{figure1} (f)). At
a large similarity degree, the scale free property (or power law
distribution) emerges, which means a few nodes have a large degree (number
of connections), but most nodes have a small degree. To make the figure \ref%
{figure1} (e) more visualized, we extracted the nodes whose connections
number is bigger than $25$ to make the figure \ref{figure2}. It also has
hubs\ in the center of the network. Of course, the hubs\ are smaller. The
scale free property is still kept.

The distribution of the connected probability of the networks of the tRNA
genes with the same anticodon is shown in table \ref{table1}. The connected
probability is defined as the fraction of number of real connections to the
largest number of possible connections. In the table it can be found, when $%
s_{0}=50$, the network is almost entirely connected and most of the
connected probabilities are larger than $0.8$; when $s_{0}=90$, most of the
connected probabilities decrease to one tenth of the former, and some
decrease to zero.

Consider the network of random tRNA sequences in the same size. When
similarity degree $s_{0}$ is small, most of the nodes have the same number
of connections. When $s_{0}$ increases, the number of the edges of the
network decreases sharply and most of the nodes lose their links; only few
of them have two or three edges linked. Table \ref{table2} shows the
statistics of the connection numbers of real tRNA similarity network and
random tRNA similarity network at different similarity degrees. The table
shows that when $s_{0}=50$, the number of the connections of the two
networks are very large; and when $s_{0}=90$, both of them drop, but the
random one drops more quickly than real one does. The connection number of
real tRNA network $n_{real}$ drops from $3434403$ ($S_{0}=50$) to $3429$ ($%
S_{0}=90$). The connection number of random tRNA network $n_{random}$ drop
from $4321688$ ($S_{0}=50$) to $0$ when $S_{0}=80$. It shows the real tRNA
sequences have more similarity with each other than random ones do. In other
words, the real tRNA sequences are not randomly taken. If we consider that
the real tRNA genes have evolutionary relationships, the differences between
the statistics of real and random tRNA similarity networks shown above can
be explained to a certain extent.

\subsection{Degree distributions}

It has already been found that networks constructed of the large scale
organization of genomic sequence segments display a transition from a
Gaussian distribution via a truncated power-law to a real power-law shaped
connectivity distribution towards increasing segment size.{\cite{Stefan
W2003}}. The similarity networks of tRNA sequences have similar features.
The investigations begin with an important parameter, degree distribution $%
p(k)$ of the nodes, and the analysis is considered in figure \ref{figure3}.

As observed in Figure \ref{figure3}, with the similarity degree $s_{0}$
increasing, $p(k)$ behaves more and more similar to power-law distribution.
When $s_{0}=50$, degree distribution $p(k)$ of the nodes follows a
uninterrupted fluctuated distribution. For those $k<1088$, $Np(k)$ fluctuate
from $1$ to $3$; and for those $k>1800$, $Np(k)$ fluctuate from $1$ to $9$,
and the peak of the fluctuation is at $k=2600$. The mean degree $\langle
k\rangle =2008$, and the maximal degree $k_{\text{max}}=3052$. When $s_{0}=60
$, the peak of the fluctuation deviates to left, at $k=100$. When $s_{0}=70$%
, the distribution of $p(k)$ appears a analogous power-law distribution if
ignore the minimal value of $k$. For $s_{0}>70$, the distribution transits
from a analogous power-law distribution to a real power-law. As shown in
figure \ref{figure3}(e), when $s=90$, the distribution curve fits the
power-law perfectly. The fitting result is $p(k)=0.192k^{-1.036}-0.006$.

Comparing to the real tRNA gene sequences, the degree distribution of the
network of random tRNA sequences, when $s_{0}=50$, is a Gaussion
distribution (figure \ref{figure3}(f)). Most nodes have approximately the
same degree, $k\thickapprox \langle k\rangle =2527$; the maximal degree $k_{%
\text{max}}=2895$ and the minimal degree $k_{\text{min}}=2327$. When $%
s_{0}=60$, the distribution is almost unchanged (figure \ref{figure3}(g)).
When $s_{0}=70$, the number of the edges descend sharply with its maximal
degree $k=5$. In figure \ref{figure3}(f), (g), there are lower peaks except
the main peaks of the Gaussion distribution. It is possibly because the
random tRNA sequences are not generated completely arbitrarily for they must
conform to the prototype of the real tRNA.

From above data analysis, we can conclude the real tRNA genes are more
self-organized than the random tRNA genes. The power-law distribution means
there are a few tRNA genes which behave as "very connected" hubs of the
similarity network. Lots of tRNA genes are similar with them in arranging of
sequences. If we suppose all the tRNA genes come from common ancestor, it is
possible that the "very connected" tRNA genes will have more relationships
with the ancestor than other tRNA genes do. In other words, the "very
connected" tRNA genes probably diverge less from ancestral sequences than
other tRNA genes do in the evolutionary history. In mathematics, a way to
construct a scale free network is to follow a rule that an added node has
much more possibility to connect with a node with a large degree than to
connect with a nod with a small degree\cite{Barabasi1999}. In the tRNA
similarity network, it maybe means the tRNA genes which have small degrees
diverged more from ancestor sequences and is less stable than the tRNA genes
which have large degrees.

\subsection{Clustering coefficient}

If a node connect with $i$ other nodes and there are $j$ edges connected
within these $i$ nodes, the clustering coefficient of the original node is
defined as%
\begin{equation*}
c=\frac{2j}{i(i-1)}
\end{equation*}
where $i\left( i-1\right) /2$ is the total number of possible connections
among $i$ nodes. Clustering coefficient reflects relationships of the
neighbors of a node, and quantifies the inherent tendency of the network to
clustering. As shown in Figure \ref{figure6}, the average clustering
coefficient $c_{\text{real}}$ of the real tRNA network is larger than the
random one. As $s_{0}$ increase, $c_{\text{real}}$ decrease. When $s_{0}=60$%
, it approaches a local minimum and experience a little increase and then
decreases slowly again. Comparing with the average clustering coefficient of
the tRNA network, the average clustering coefficient $c_{\text{random}}$ of
the random network decreases fast while $s_{0}$ increases, when $s_{0}>70$, $%
c_{\text{random}}\rightarrow 0$. The behavior of the coefficient of two
networks is also illustrated in table \ref{table2}. When $s_{0}=50$, $c_{%
\text{real}}=0.777367$, $c_{\text{random}}=0.747479$; when $s>70$, $c_{\text{%
random}}$ drops to zero quickly, but $c_{\text{real}}$ decrease slowly. Once
again, we proved the real tRNA genes are not randomly selected. The real
tRNA genes have close relationships with each other.

Table \ref{table3} shows the distribution of the average clustering
coefficient of $19$ tRNA groups which are classfied by the possible amino
acid-accepting. Some groups contain isoacceptor tRNA which consist of
different tRNA species that bind to alternate codons for the same amino acid
residue. The tRNA group who carries the amino acid residue named Met is
ignored for it contains only one tRNA sequence. Comparing table \ref{table3}
with table \ref{table2}, we can conclude that the nodes are more likely to
connect with the nodes within the same amino acid group. The tRNA similarity
network can be classified into several large clusters with the same amino
acids. It hints that in tRNA genes evolutionary step is much more likely to
happen within the same amino acid group. The cases that a tRNA gene of
certain amino acid evolve to tRNA gene of another amino acid are rare.

\section{Discussion}

\bigskip In this paper, we want to show the network model is a powerful tool
to study the relationship of tRNA genes. Although some results are not new,
such as the real tRNA genes are not random and the relationships among tRNA
genes with same anticodon are closer than the relationships among tRNA genes
with different anticodons, they are evidences that network model works well
for the network model distinguishes these properties clearly. What is more,
the tRNA similarity network behaves scale-free properties when $s_{0}$ is
large. As we know the scale-free nature is rooted in two generic mechanisms%
\cite{Barabasi1999}. Firstly scale-free networks describe open systems that
grow by the continuous addition of new nodes. Secondly scale-free networks
exhibit preferential attachment that means the likelihood of connecting to a
node depends on the node's degree. With these mechanisms, the "very
connected" nodes in scale-free networks usually are added in the network at early time
during the growth of the network. It has been found that most recent tRNA
genes are evolved from a few common precursors{\cite{Sergey1993, Robert1999}}%
, and these oldest evolutionary sequences, comparing to the recent tRNA
genes. Therefore, in tRNA similarity netwok, the "very connected" tRNA genes
may have diverged less from their ancestors than weakly connected ones.

Most recently, many research conclusions show that genes of related function
could behave together as a group in the networks constructed according to
their similarity features{\cite{Himanshu2002, Stefan2001}}. In this paper,
although we use the simplest alignment, this property can be found. When
similarity degree $s_{0}$ is small, nodes of the tRNA genes with the same
anticodons are connected to form a local cluster, among them are entirely
connected. When $s_{0}$ increases to a large value, a scale-free character
emerges that a few nodes compose the core of the network and most of nodes
have low links. These observations seem to be perfectly fit to the
evolutionary processes of the tRNA genes. On the other hand, the oldest tRNA
genes undergo disturbances such as mutation, loss, insertion, or
rearrangement etc. during the evolution. Some new tRNA genes are suited for
the environment and reserved. So, they have a high similarity to its
ancestral sequences. In the network constructed by similarity degree of
these tRNA genes, they form local clusters.

An interesting finding of tRNA similarity networks is that some local
clusters have high connectivity with the other clusters; or to say, some
nodes of one cluster have lots of connections with some nodes of another
cluster. See figure \ref{figure5}. It may hint that the evolution
relationship of tRNA sequences of two different anitcodons. As shown in
figure\ref{figure5}(a), the network is of two different anticodons: ACG and
CCA. The solid circle nodes are the tRNA genes of ACG, and the hollow circle
nodes are the tRNA genes of CCA. In this figure, they mix into one cluster.
Figure \ref{figure5}(b) shows that the network of anticodons TAG and TGA.
The solid circle nodes are the tRNA genes of TAG, and the hollow circle
nodes are tRNA genes of TGA. They appear three clusters in the topology map,
and each cluster has some nodes which highly connect with some nodes of
other clusters. It shows that although some tRNA genes have different
anticodons, they have high similarities in sequences. In evolutionary
history, the tRNA genes of one anticodon identity can evolve to tRNA genes
of another identity. The above finding may be an evidence of this kind of
evolutionary mode. In the other hand, from figure \ref{figure1}(c), the
network of the same anticodon GCC split into two cluster. It hints the
evolution process of the tRNA genes of same anticodon may diverge in the
history. Therefore, there are different modes of evolutionary processes,
i.e. evolution within the same anticodon groups and evolution among
different anticodon groups. The former may be the main part of tRNA
evolution. The later may be the key cases of the interaction among tRNA of
different anticodons during the evolution.

For the alignment we used is simply counting the number of cites that are
identical, it losts many information in the evolution process. More
complicated alignment models may exhibit more details of the relationships
among tRNA genes. The content of tRNA database is limited, the numbers of
tRNA sequences from different organisms varied largely. Therefore, the
biases of taxon samples may influence the topology of the network and the
results gotten from the network may not completely reflect the evolution
relationship of tRNA genes. It is a limitation of network model that will be
improved when more tRNA genes are sequenced. Although we did not get many
new results from what we have know about the evolution of tRNA genes, the
results contribute as proofs that the network model can work well in the
research of relationship of tRNA genes and is a useful tool.

\begin{acknowledgments}
This paper was supported by the National Science Foundation of China under
Grant No. 10105007, No. 10334020 and No. 90103035. .
\end{acknowledgments}

\newpage

\renewcommand{\baselinestretch}{1.3} {\normalsize
\begin{table}[tbp]
\begin{center}
{\normalsize
\begin{tabular}{cccccc||c||cccccc}
\hline\hline
$s_0$ & 50 & 60 & 70 & 80 & 90 &  & gat & 0.8508 & 0.4392 & 0.2912 & 0.1326
& 0.0389 \\
aac & 0.8182 & 0.8182 & 0.7636 & 0.3636 & 0.1091 &  & gca & 0.8311 & 0.4082
& 0.175 & 0.0492 & 0.0131 \\
aag & 0.8 & 0.6 & 0.6 & 0.1 & 0 &  & gcc & 0.9832 & 0.8225 & 0.5423 & 0.1964
& 0.0461 \\
aat & 1 & 0.9333 & 0.7333 & 0.1333 & 0.0667 &  & gcg & 1 & 1 & 0.3333 &
0.0667 & 0 \\
acg & 0.9111 & 0.8537 & 0.4378 & 0.1347 & 0.0604 &  & gct & 0.4509 & 0.1226
& 0.1226 & 0.0582 & 0.0076 \\
agc & 1 & 1 & 0.9615 & 0.5513 & 0.2692 &  & gga & 0.9415 & 0.7969 & 0.4092 &
0.1662 & 0.0738 \\
agg & 1 & 0.7333 & 0.6667 & 0.5333 & 0.0667 &  & ggc & 1 & 0.8095 & 0.8095 &
0.3333 & 0.0476 \\
agt & 0.75 & 0.7142 & 0.3929 & 0.0357 & 0.0357 &  & ggg & 1 & 0.9722 & 0.5278
& 0.1944 & 0 \\
acc & 1 & 0.7 & 0.2 & 0.1 & 0.1 &  & gta & 0.7257 & 0.3637 & 0.1621 & 0.054
& 0.0153 \\
act & 1 & 1 & 1 & 1 & 0 &  & gtc & 0.8347 & 0.4392 & 0.1849 & 0.0643 & 0.0171
\\
aga & 1 & 1 & 0.7867 & 0.4338 & 0.1691 &  & gtg & 0.9293 & 0.5676 & 0.2688 &
0.1367 & 0.0262 \\
caa & 0.8842 & 0.6182 & 0.2192 & 0.1035 & 0.0025 &  & gtt & 0.9349 & 0.6379
& 0.28 & 0.0899 & 0.0214 \\
cac & 0.9455 & 0.6909 & 0.4727 & 0.2364 & 0.1273 &  & ggt & 0.8996 & 0.8655
& 0.6231 & 0.1989 & 0.0417 \\
cag & 0.7493 & 0.4431 & 0.1396 & 0.06268 & 0.0256 &  & taa & 0.6036 & 0.312
& 0.1466 & 0.0652 & 0.0185 \\
cat & 0.9034 & 0.486 & 0.1704 & 0.0549 & 0.0194 &  & tac & 0.9138 & 0.5152 &
0.2036 & 0.0496 & 0.0111 \\
cca & 0.9874 & 0.8414 & 0.3676 & 0.0985 & 0.037 &  & tag & 0.6542 & 0.4008 &
0.2711 & 0.2153 & 0.0584 \\
ccc & 0.8611 & 0.7778 & 0.4722 & 0.0833 & 0 &  & tat & 0.8 & 0.6 & 0.6 & 0 &
0 \\
ccg & 0.8182 & 0.8181 & 0.3455 & 0.1636 & 0.1091 &  & tca & 0.8318 & 0.6033
& 0.2827 & 0.0623 & 0.0079 \\
cct & 1 & 0.9091 & 0.3818 & 0.0364 & 0 &  & tcc & 0.9142 & 0.503 & 0.221 &
0.0573 & 0.0093 \\
cga & 0.9083 & 0.45 & 0.1167 & 0.0417 & 0.0167 &  & tcg & 0.8136 & 0.4711 &
0.2653 & 0.1251 & 0.0152 \\
cgc & 1 & 1 & 0.8667 & 0.0667 & 0 &  & tct & 0.8865 & 0.7358 & 0.2385 &
0.0621 & 0.03014 \\
cgg & 1 & 1 & 0.6667 & 0.3809 & 0.0476 &  & tga & 0.5373 & 0.286 & 0.141 &
0.0443 & 0.0066 \\
cgt & 1 & 0.956 & 0.4725 & 0.0549 & 0.011 &  & tgc & 0.9195 & 0.5892 & 0.339
& 0.1261 & 0.0257 \\
ctc & 1 & 0.9809 & 0.6 & 0.2857 & 0.0762 &  & tgg & 0.8865 & 0.5023 & 0.1899
& 0.0566 & 0.0121 \\
ctg & 1 & 0.8053 & 0.3474 & 0.1211 & 0.0947 &  & tgt & 0.9493 & 0.59 & 0.2561
& 0.0529 & 0.0081 \\
ctt & 0.7808 & 0.5045 & 0.3649 & 0.1156 & 0.036 &  & tta & 1 & 1 & 0 & 0 & 0
\\
gaa & 0.9424 & 0.5709 & 0.2778 & 0.0891 & 0.0151 &  & ttc & 0.8858 & 0.4864
& 0.2131 & 0.0728 & 0.0165 \\
gac & 1 & 1 & 0.4314 & 0.2026 & 0.1503 &  & ttg & 0.9544 & 0.5946 & 0.2116 &
0.0642 & 0.0098 \\
gag & 0.9636 & 0.5818 & 0.2909 & 0.0545 & 0 &  & ttt & 0.9119 & 0.5172 &
0.2613 & 0.0613 & 0.0146 \\ \hline\hline
\end{tabular}
}
\end{center}
\par
{\normalsize 
}
\caption{The distribution of the connected probability of all 57 anticodons'
tRNAs networks, which have excluded four anticodons for they have too small
vertices. The statistic shows that when s=50, many networks are complete
connection; when $s_0$=90, the connected probability decreasing sharply,
some of the connected probability decrease to zero}
\label{table1}
\end{table}
}

{\normalsize \newpage }

{\normalsize
\begin{table}[tbp]
\begin{center}
{\normalsize
\begin{tabular}{cllll}
\hline\hline
$s_0$ & $n_{real}$ & $n_{random}$ & $c_{real}$ & $c_{random}$ \\ \hline
50 & 3434403 & 4321688 & 0.777367 & 0.747479 \\
60 & 994571 & 367845 & 0.541708 & 0.139572 \\
70 & 142264 & 773 & 0.578806 & 0.000682 \\
80 & 19453 & 0 & 0.567380 & 0.000000 \\
90 & 4249 & 0 & 0.286254 & 0.000000 \\ \hline\hline
\end{tabular}
}
\end{center}
\caption{The number of edges and average cluster coefficients of two
networks respective to similarity degrees. The number of nodes is 3420. $S_0$%
: similarity degree; $n$: the number of edges of the network; $c$: average
cluster coefficient}
\label{table2}
\end{table}
}


{\normalsize
\begin{table}[tbp]
\begin{center}
{\normalsize
\begin{tabular}{clllll}
\hline\hline
$s_0$ & 50 & 60 & 70 & 80 & 90 \\
VAL & 0.838031 & 0.653432 & 0.704976 & 0.61135 & 0.302831 \\
ASH & 0.868872 & 0.878484 & 0.745679 & 0.306494 & 0.151515 \\
ASP & 0.924069 & 0.814945 & 0.818599 & 0.637089 & 0.391828 \\
CYS & 0.928983 & 0.765042 & 0.717865 & 0.637071 & 0.320786 \\
ALA & 0.738776 & 0.745374 & 0.6728 & 0.481527 & 0.253499 \\
GLN & 0.865414 & 0.727525 & 0.630352 & 0.338557 & 0.245455 \\
GLU & 0.929629 & 0.808843 & 0.767439 & 0.559427 & 0.260123 \\
GLY & 0.888469 & 0.874198 & 0.683965 & 0.540341 & 0.200635 \\
HIS & 0.93883 & 0.745659 & 0.716469 & 0.677458 & 0.406243 \\
LEU & 0.921699 & 0.681914 & 0.733301 & 0.609716 & 0.313292 \\
LYS & 0.722222 & 0.666667 & 0.666667 & 0 & 0 \\
PHE & 0.877036 & 0.678511 & 0.744765 & 0.621813 & 0.287326 \\
PRO & 0.97332 & 0.848348 & 0.546192 & 0.447778 & 0.18366 \\
SER & 0.856441 & 0.666374 & 0.649365 & 0.542714 & 0.175925 \\
STOP & 0.758638 & 0.707049 & 0.731592 & 0.57147 & 0.217136 \\
THR & 0.93964 & 0.831404 & 0.6121 & 0.585127 & 0.309603 \\
TRP & 0.9265 & 0.789019 & 0.754142 & 0.581435 & 0.225642 \\
TYR & 0.932773 & 0.779707 & 0.684642 & 0.558649 & 0.306018 \\
ARG & 0.805026 & 0.834632 & 0.632049 & 0.375894 & 0.147186 \\
&  &  &  &  &
\end{tabular}
}
\end{center}
\caption{The average clustering coefficient of 19 tRNA possible
aminoacid-accepting groups' networks, each network is named using
three-letter amino acid abbreviations.}
\label{table3}
\end{table}
}

{\normalsize \newpage }

{\normalsize
\begin{figure}[tbp]
{\normalsize \epsfig{file=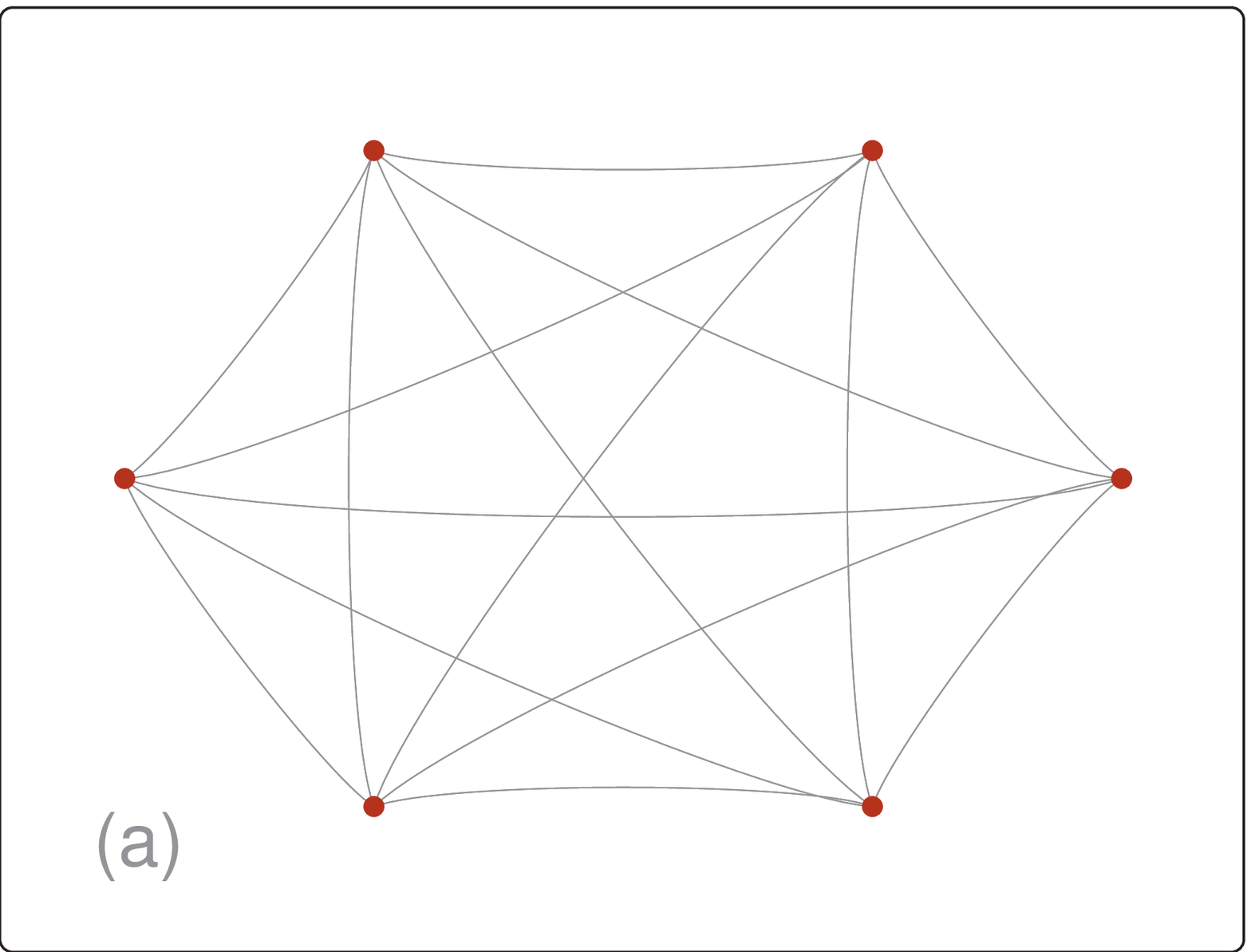, width=0.45\linewidth}
\epsfig{file=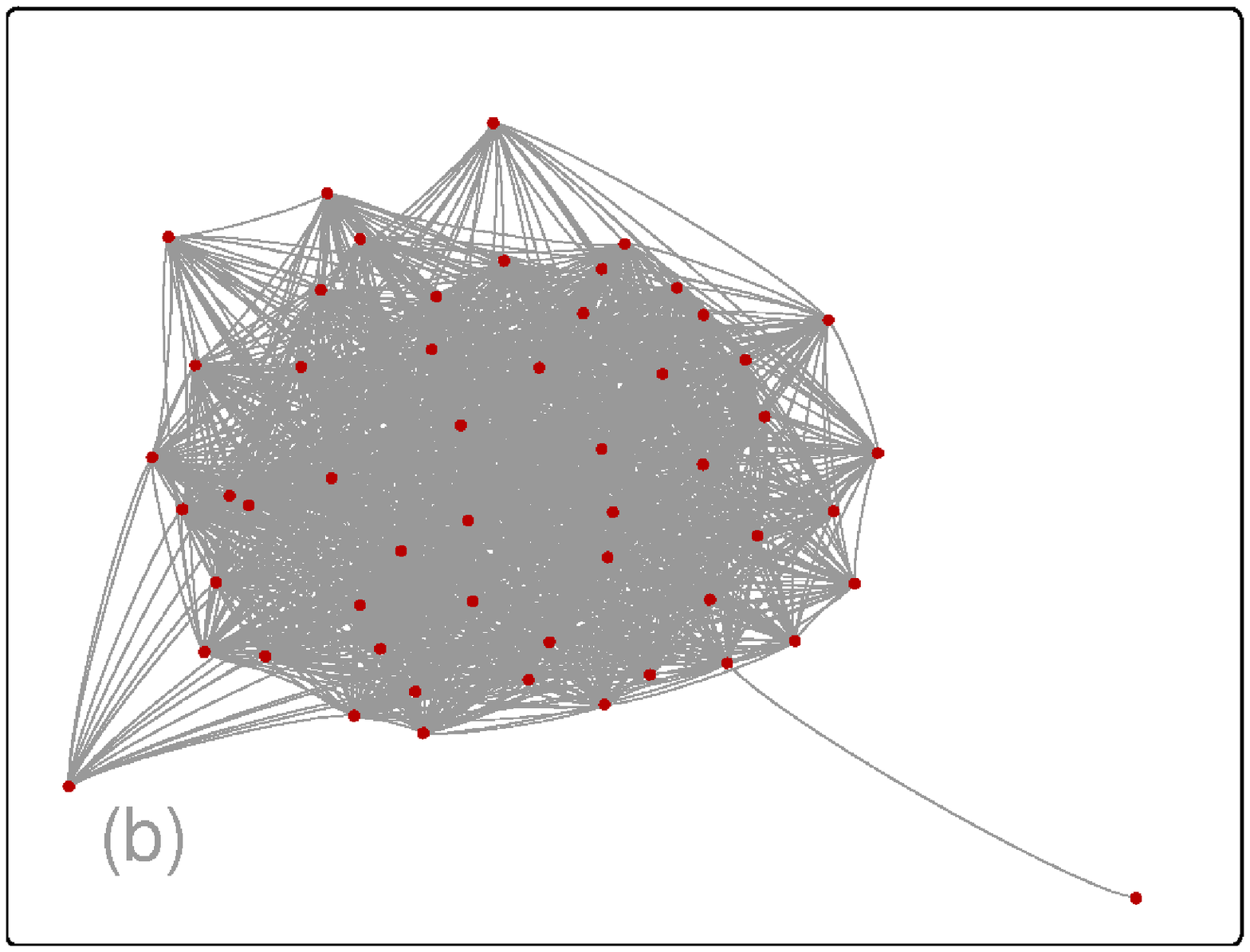,
width=0.45\linewidth} \newline
\newline
\epsfig{file=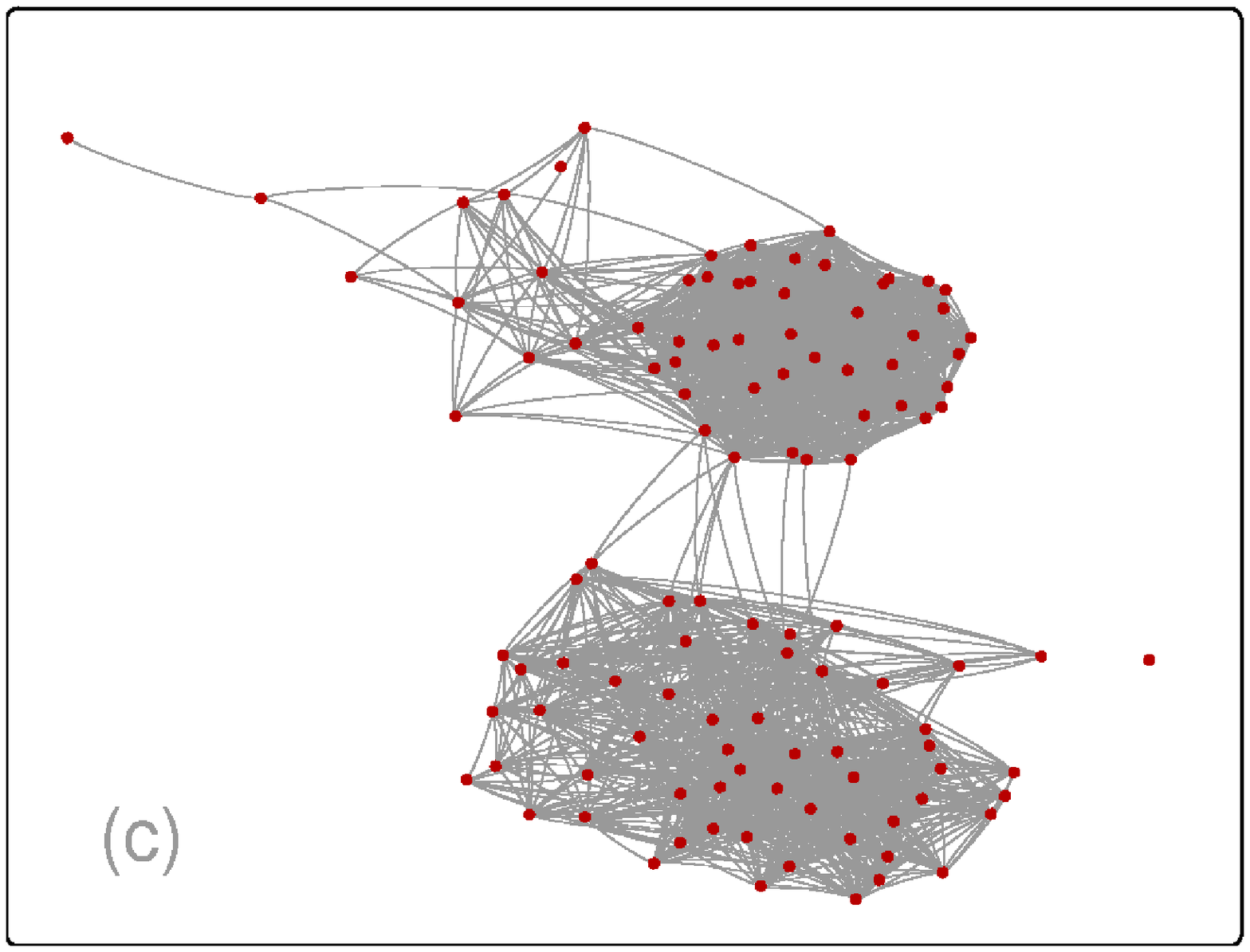, width=0.45\linewidth} \epsfig{file=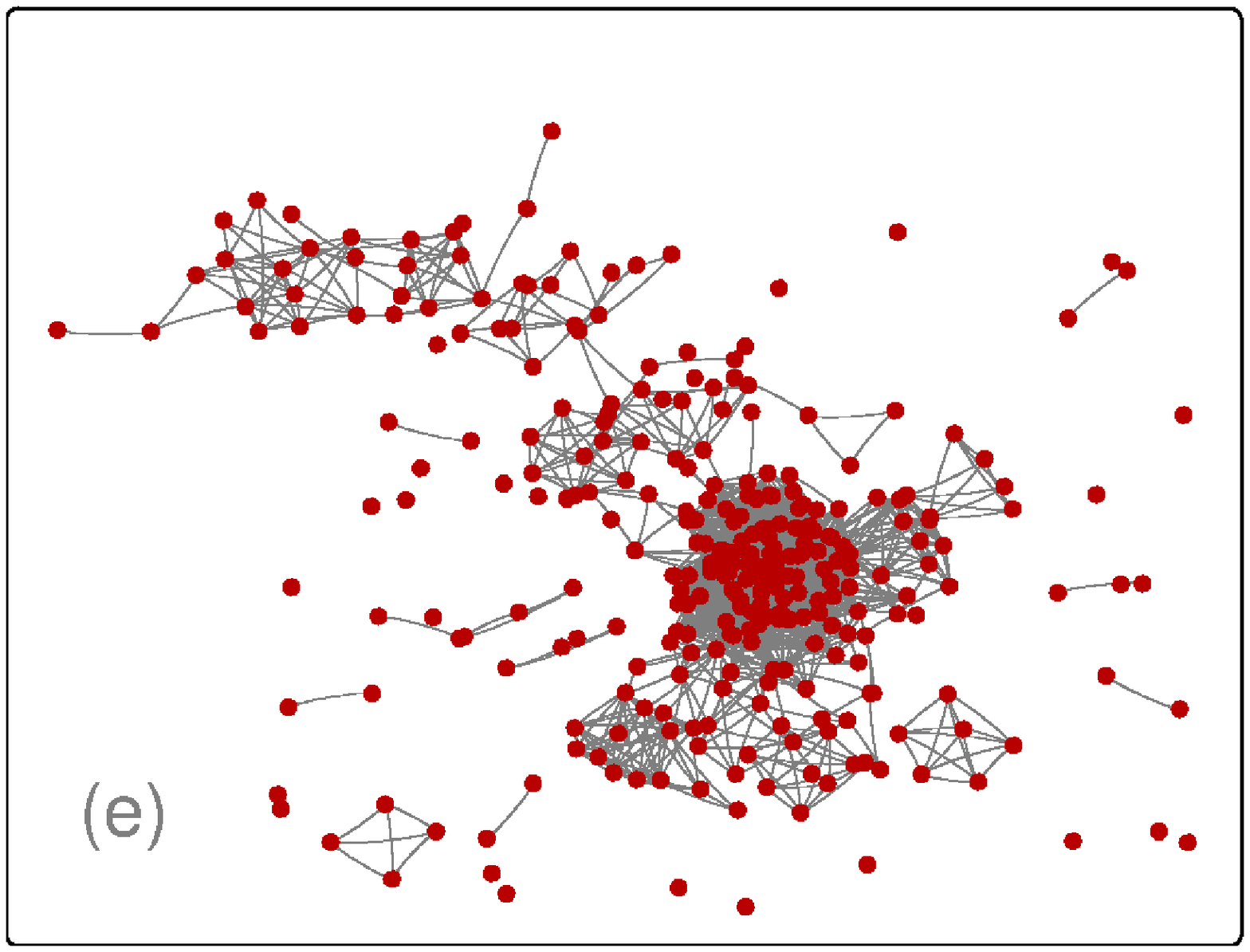,
width=0.45\linewidth,height=2.2in} \newline
\newline
\epsfig{file=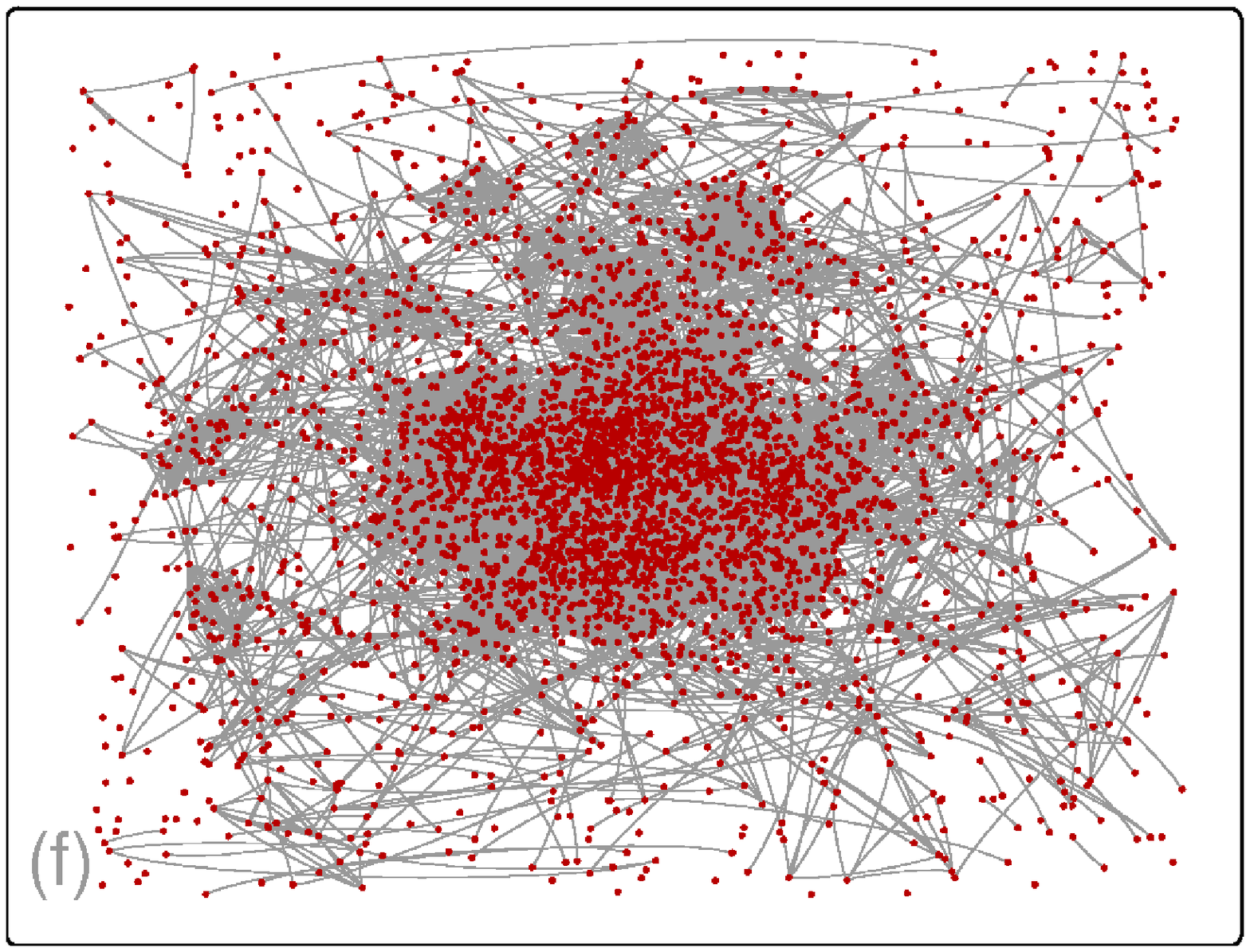, width=0.45\linewidth} \epsfig{file=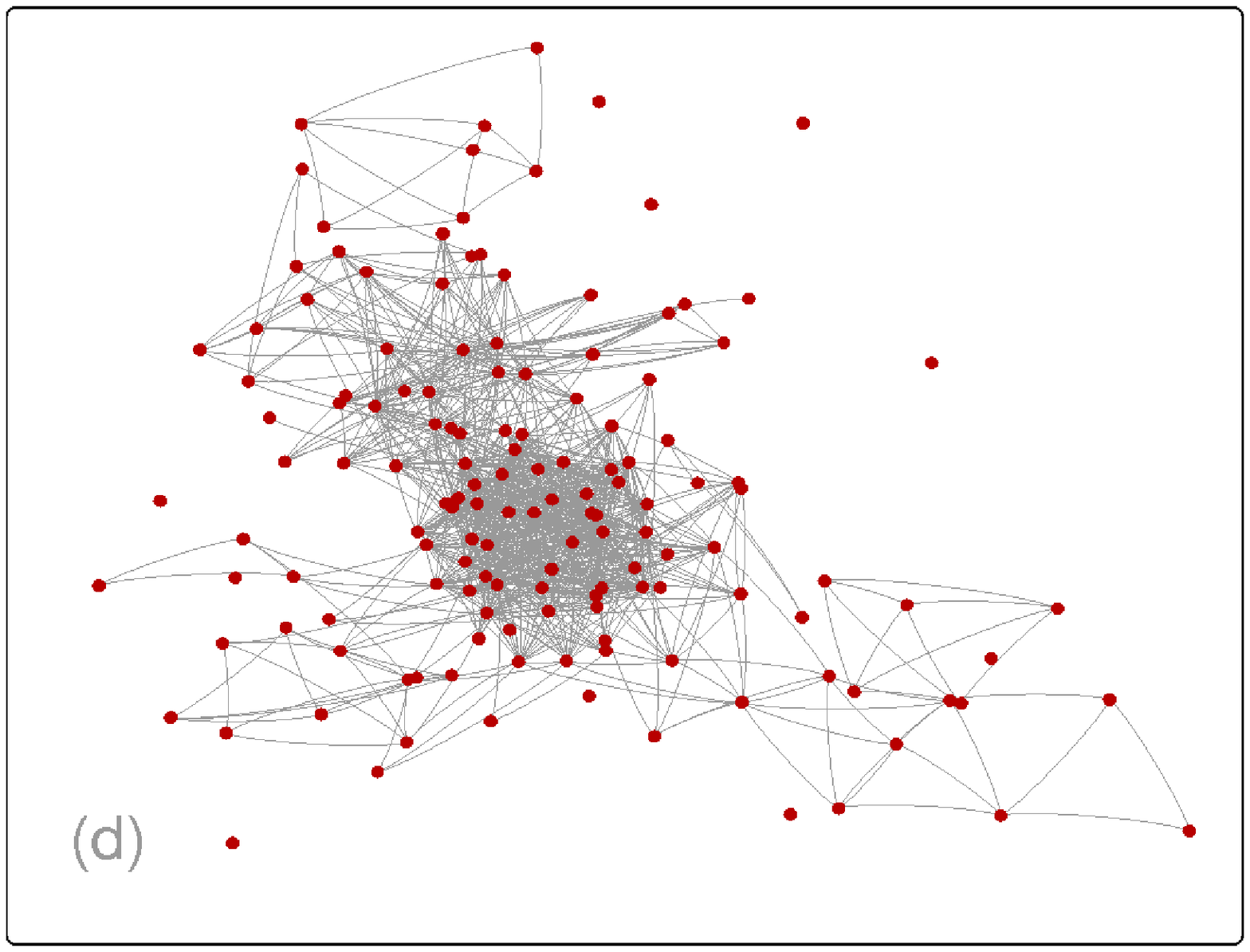,
width=0.45\linewidth} \newline
}
\caption{the topology of the network. (a), (b), (c), (d)) are the topology
of network of the same anticodons. The three capital letters are the three
anticodons subsets of tRNA genes,(a): CGC, $S_0=60,\,N=6,\,P=1.0$; (b): CCA,$%
S_0=60,\,N=150,\,P=0.8414$; (c): TGC $S_0=60,\,N=215,\,P=0.5892$ (d): GTT $%
S_0=80,\,N=145,\,P=0.028$)). (e), (f) are the topology of network of
different anticodons. (e): $S_0=80,\,N=304,\,P=0.04$, network of CAT and
GCC; (f): $S_0=90,\,N=3420,\,P=0.0034$. $S_0$ is the similarity degree; $N$
is the number of the nodes; $P$ is the connection probability. }
\label{figure1}
\end{figure}
}

{\normalsize \newpage }

{\normalsize
\begin{figure}[tbp]
{\normalsize \epsfig{file=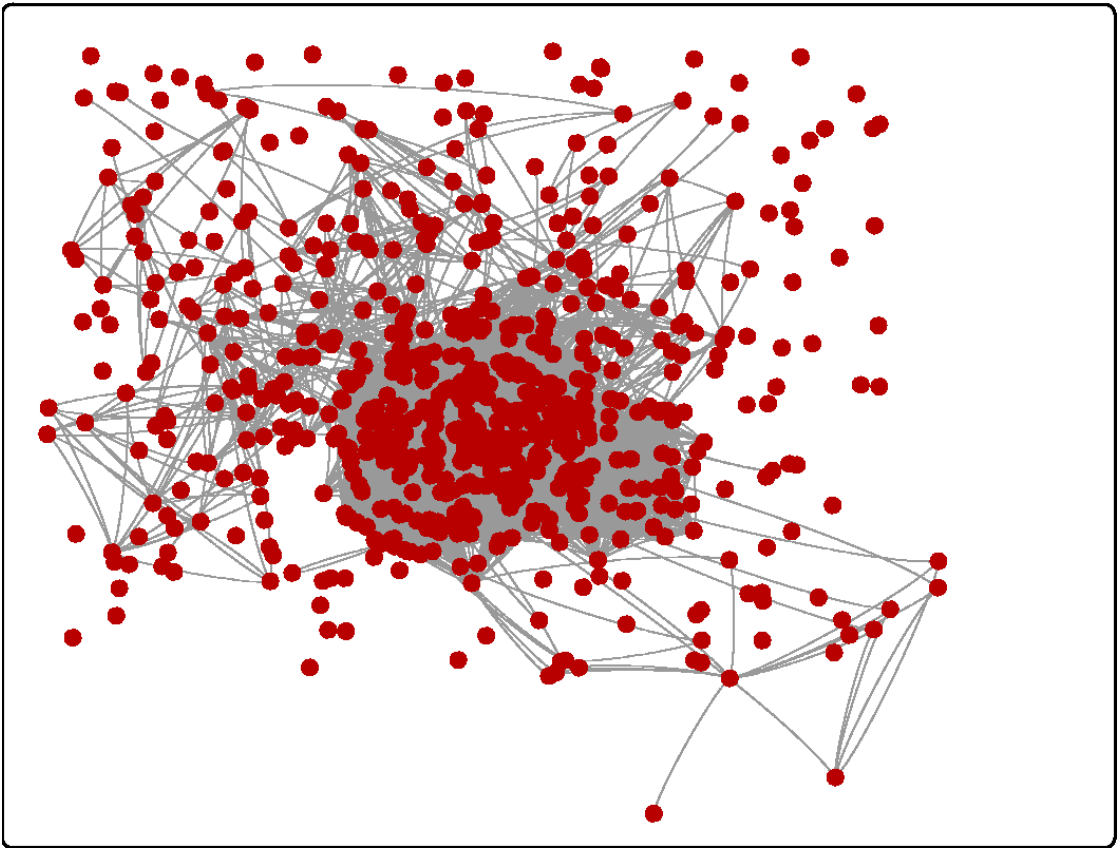,
width=0.90\linewidth}  }
\caption{The topology of network which extract the nodes which degree $%
k\geq25$ from figure 1 (f). $S_0=90$}
\label{figure2}
\end{figure}
}

{\normalsize \newpage }

{\normalsize
\begin{figure}[tbp]
{\normalsize \epsfig{file=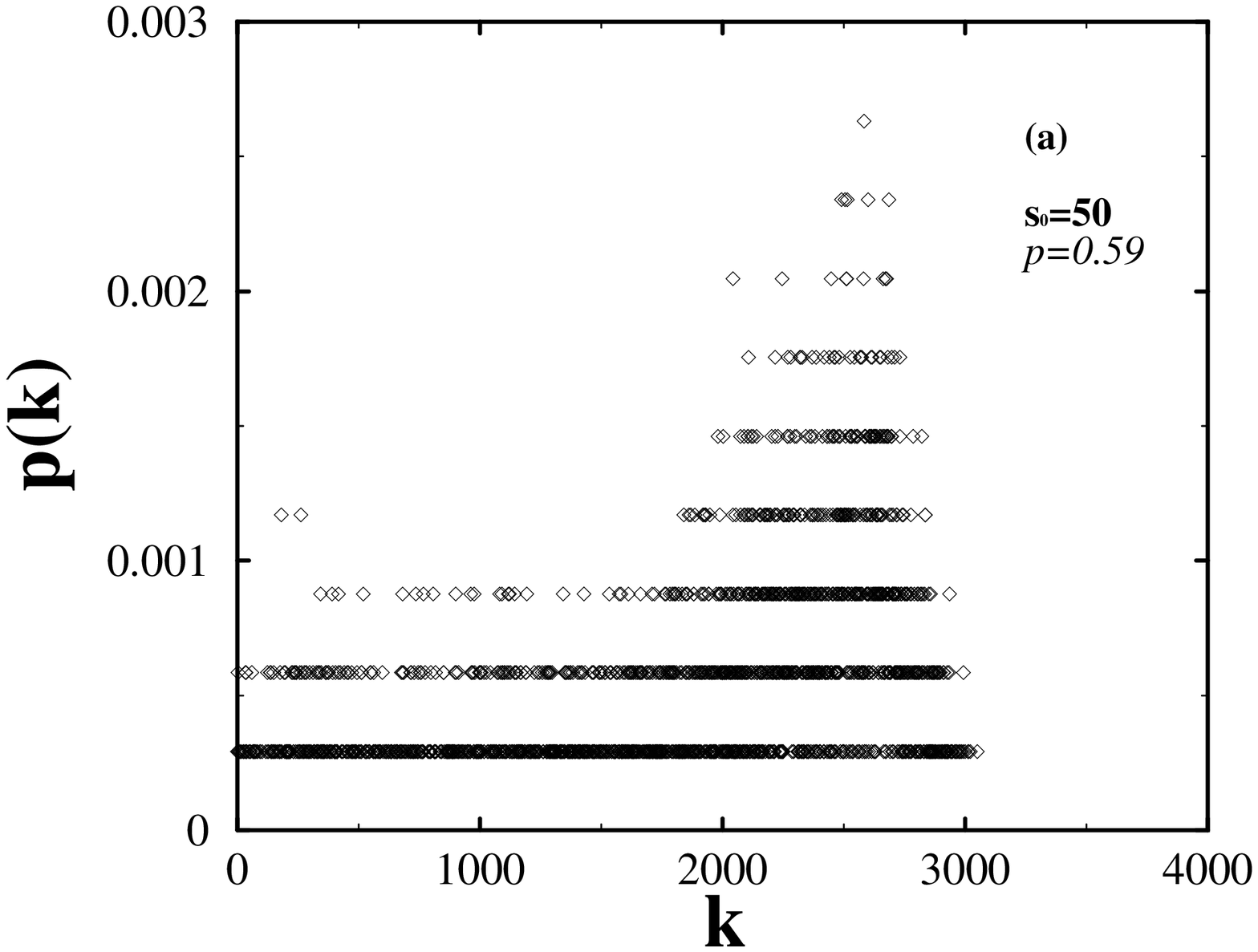,
width=3in,
height=2.1in} \,\,\,\,\, \epsfig{file=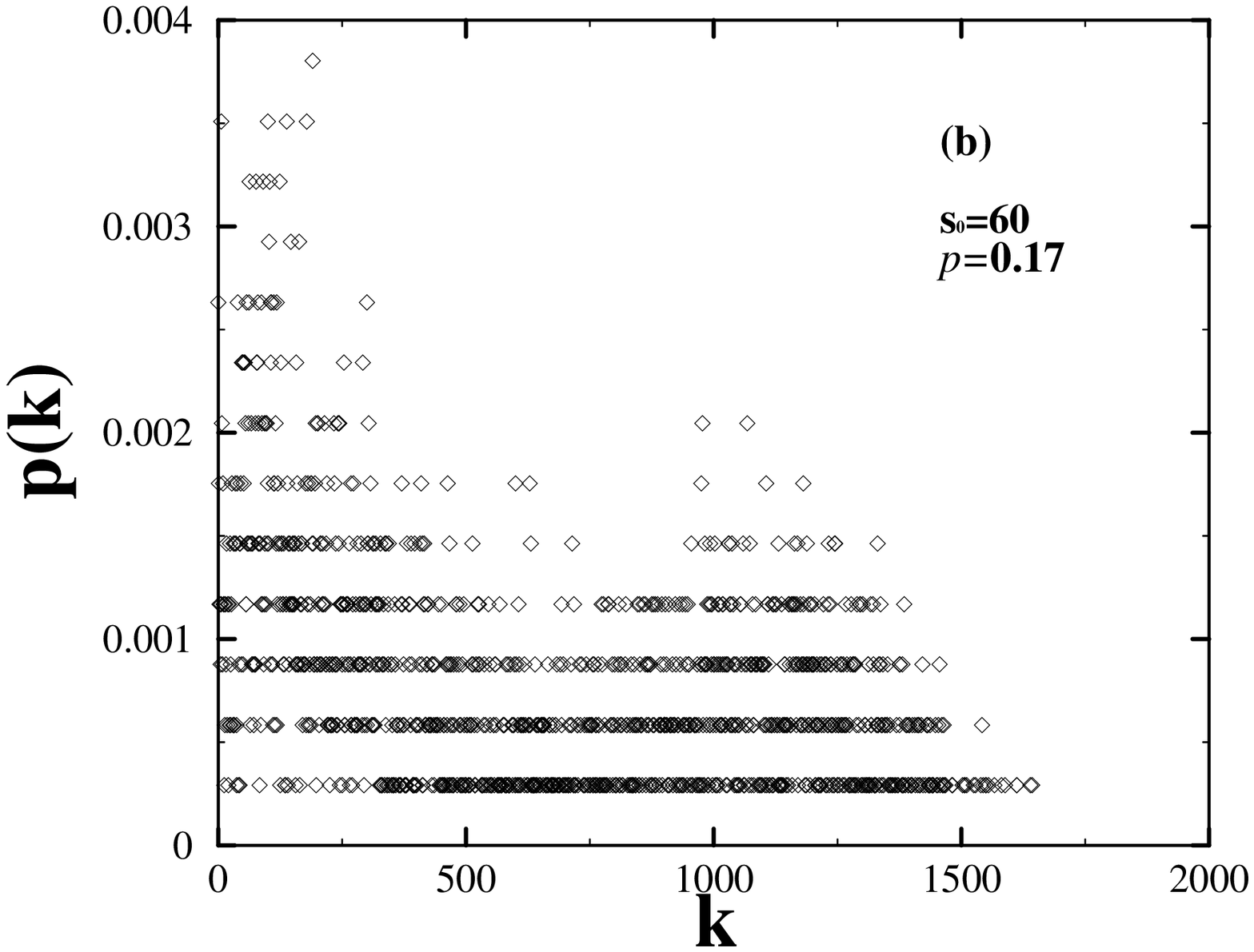, width=3in, height=2.1in}
\newline
\epsfig{file=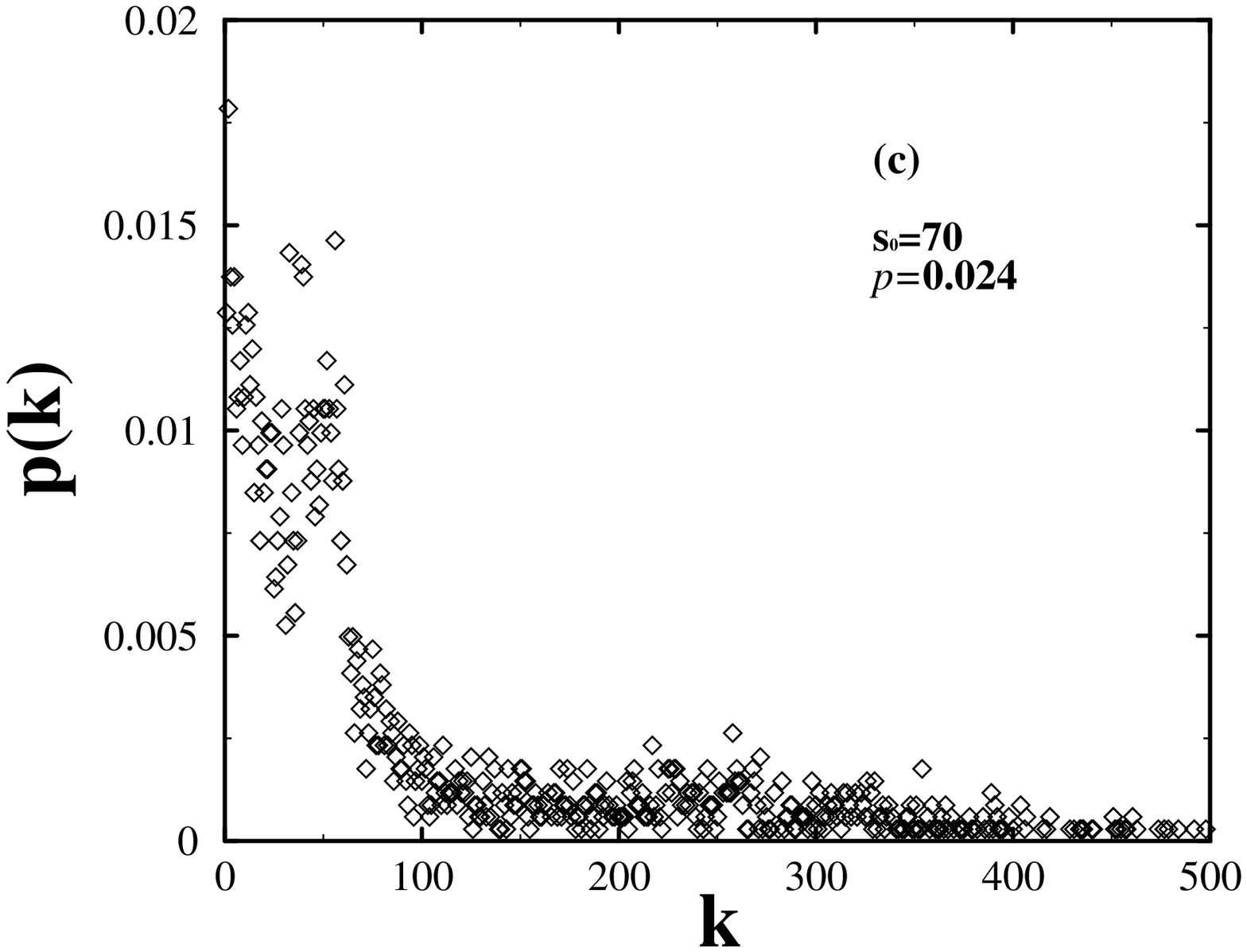,
width=3in, height=2.1in}\,\,\,\,\,  \epsfig{file=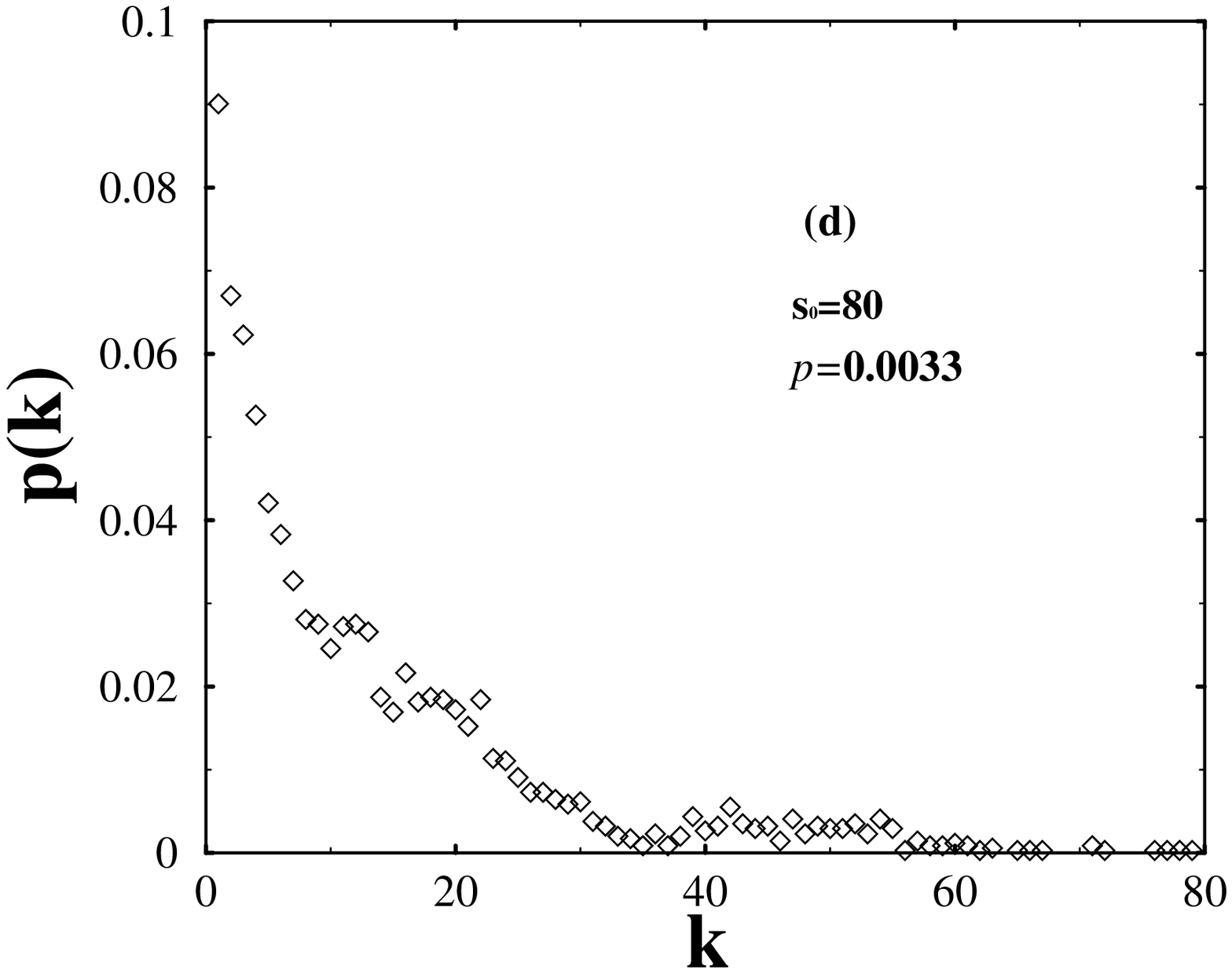, width=3in,
height=2.1in}  \newline
\epsfig{file=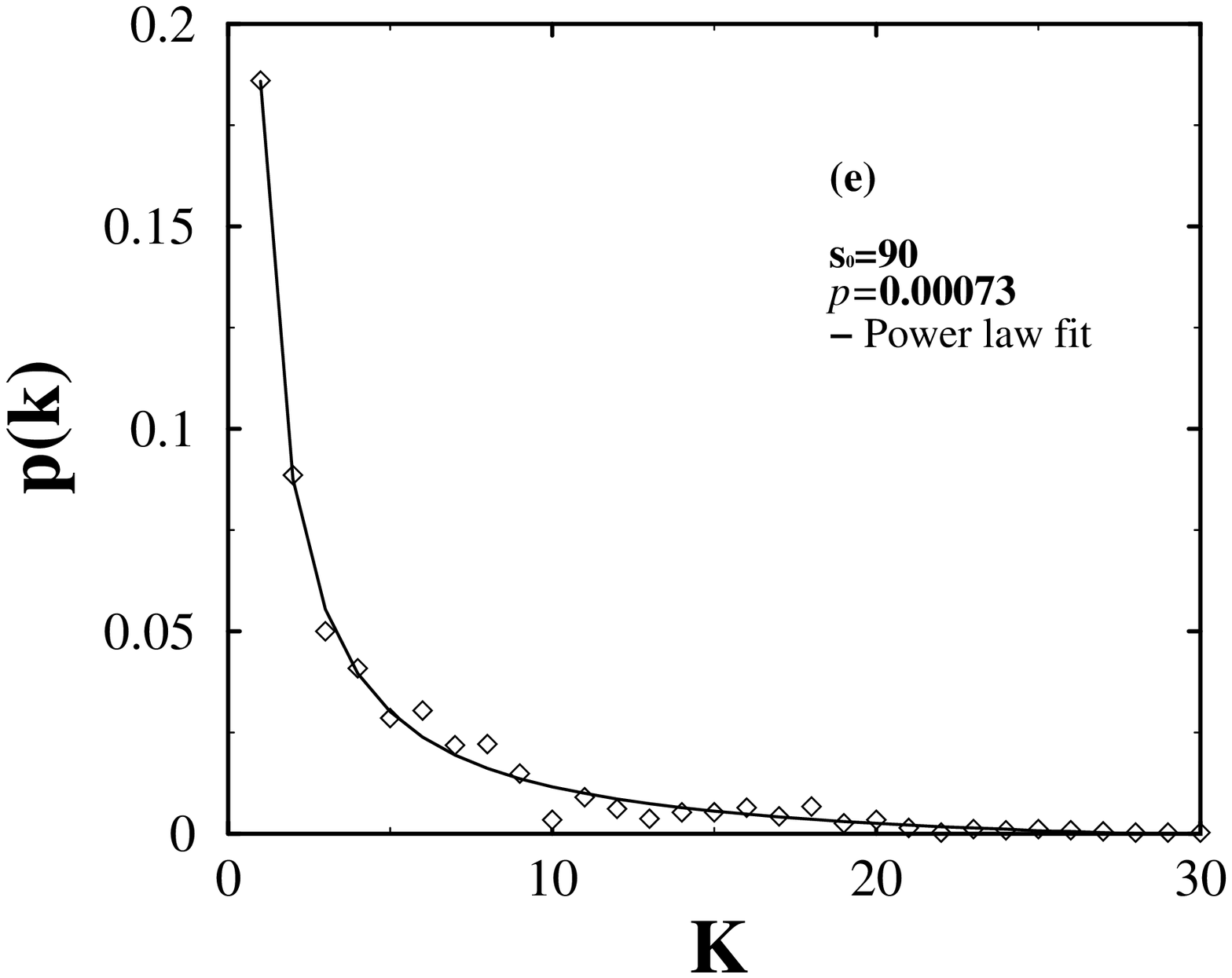, width=3in, height=2.1in} \,\,\,\,\, %
\epsfig{file=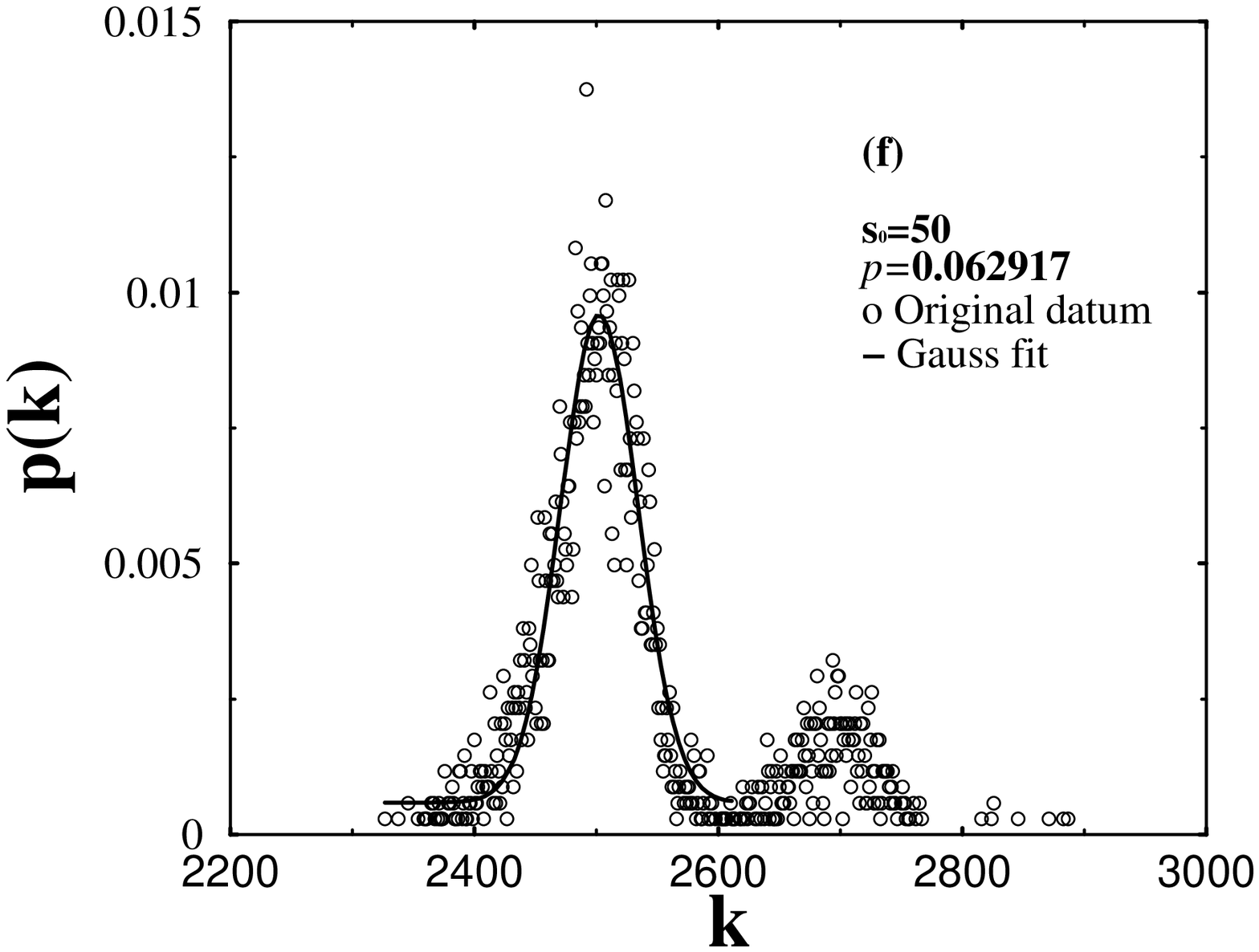,width=3in, height=2.1in} \newline
\epsfig{file=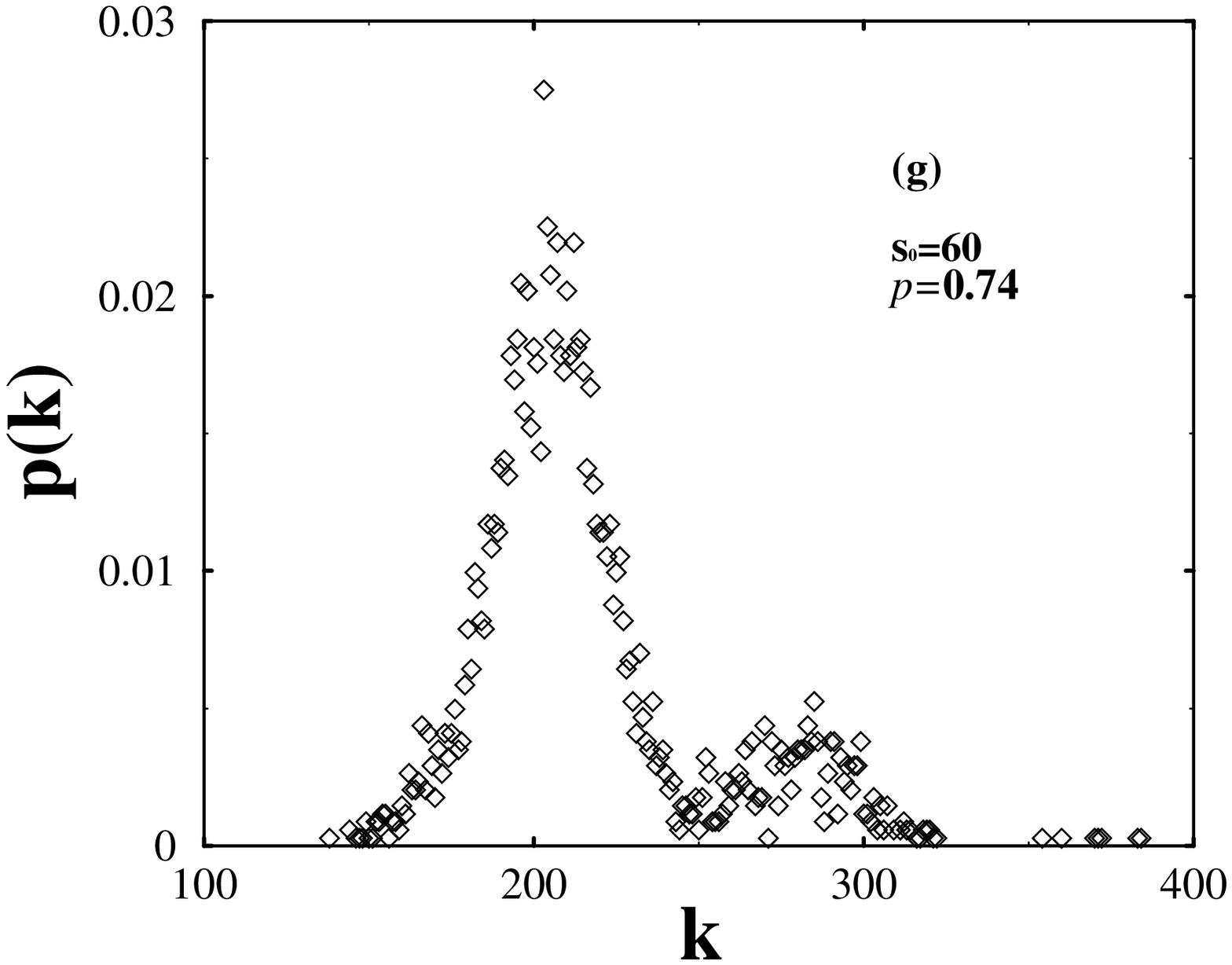,
width=3in,
height=2.1in}  }
\caption{(a), (b), (c), (d), (e)are the degree distribution of the tRNA gene
sequences network, N=3420; The line in (e) is power law fitting of the data.
The formula is $p(k)=0.192k^{-1.036}-0.006$. (f), (g) are the degree
distribution of the random tRNA sequence network, N=3420.}
\label{figure3}
\end{figure}
}

{\normalsize \newpage }

{\normalsize
\begin{figure}[tbp]
{\normalsize \epsfig{file=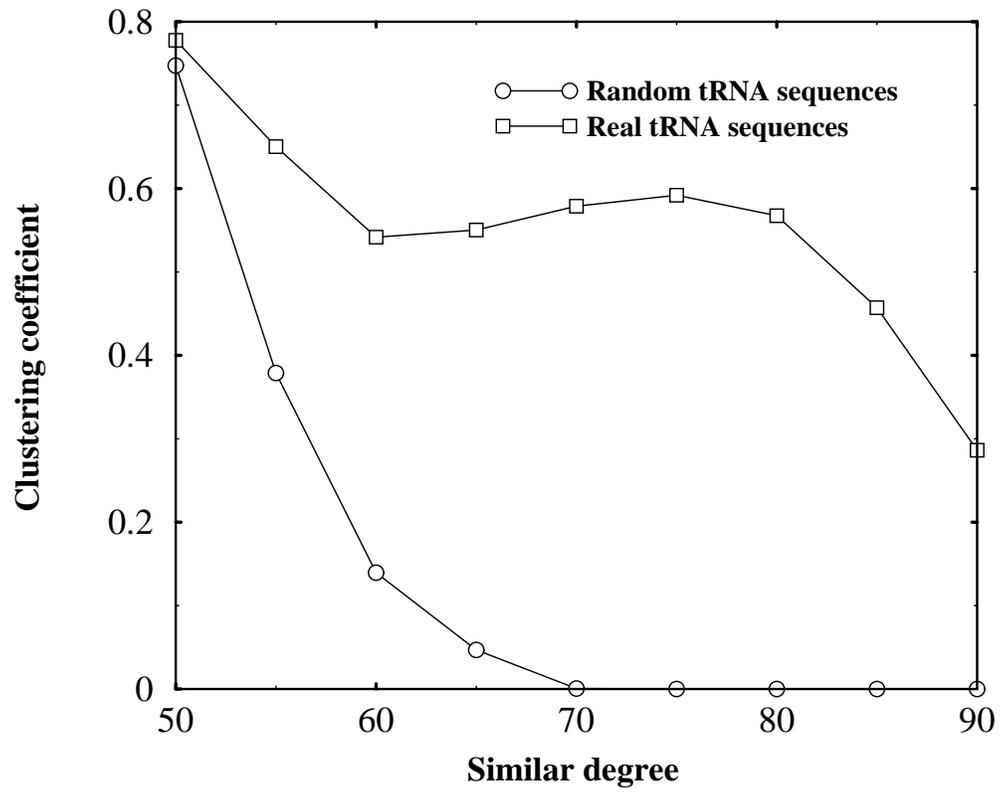, width=0.80\linewidth}  }
\caption{The distribution of the clustering coefficient of the two network
according to their similarity degree}
\label{figure6}
\end{figure}
}

{\normalsize \newpage }

{\normalsize
\begin{figure}[tbp]
{\normalsize \epsfig{file=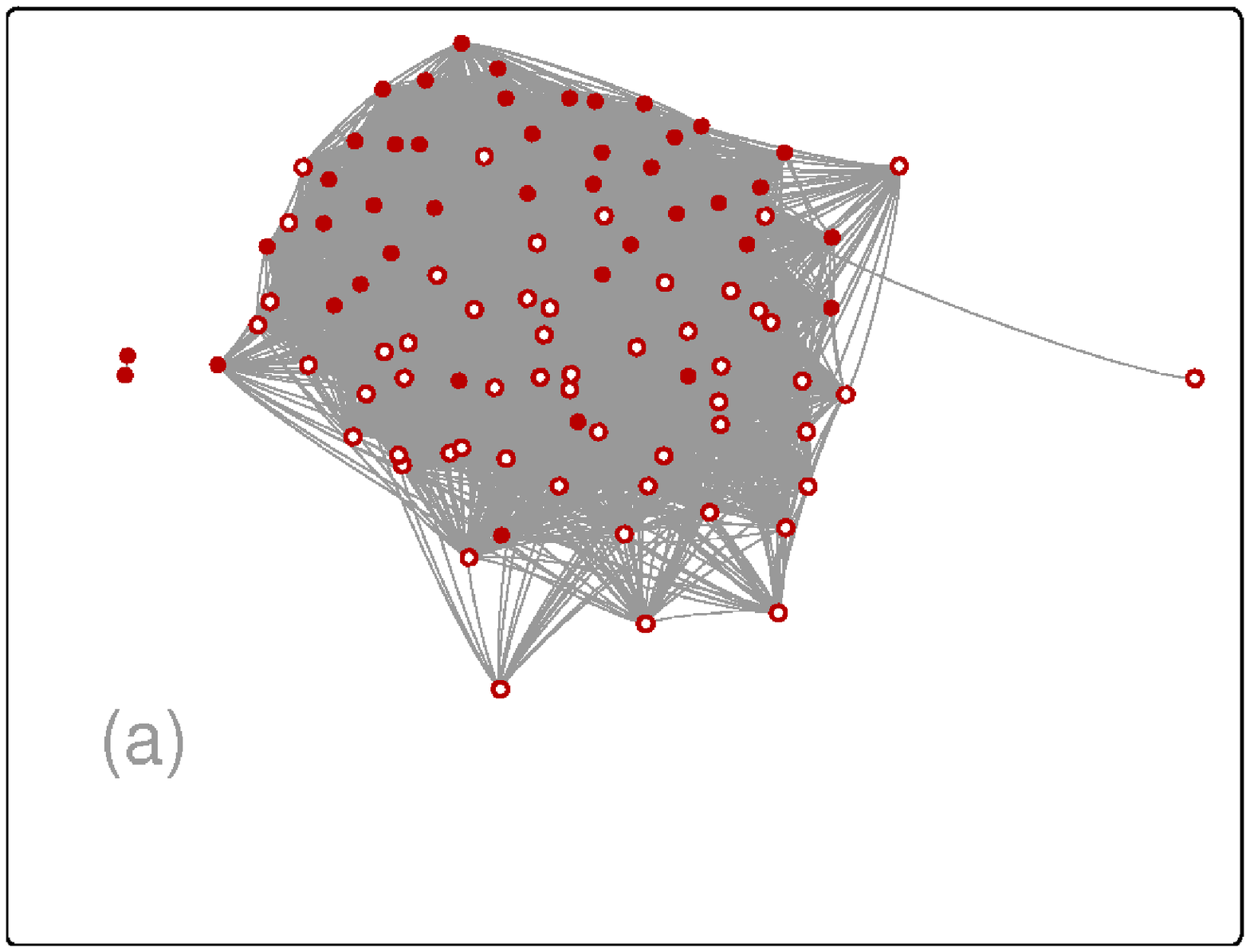, width=0.80\linewidth} %
\epsfig{file=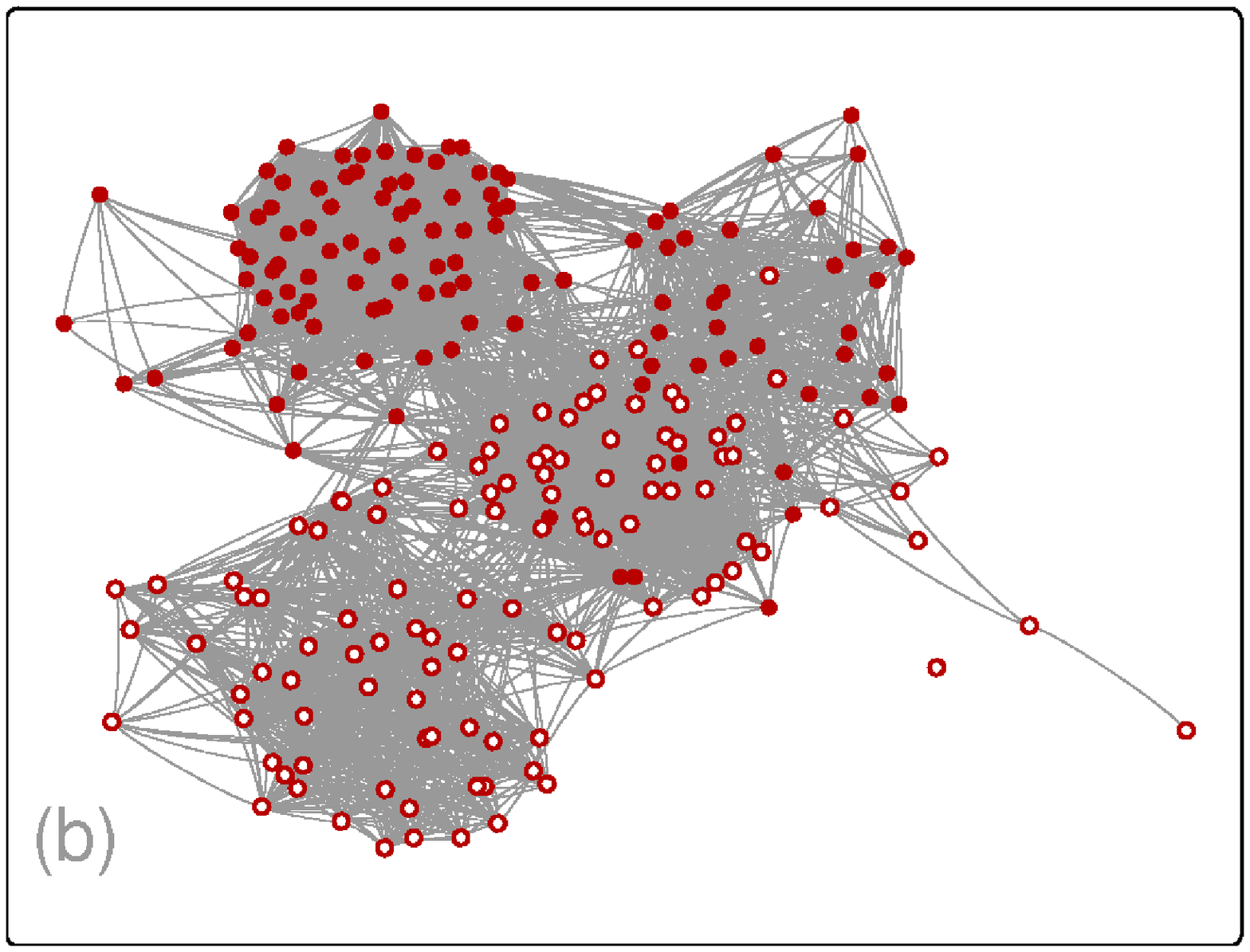, width=0.80\linewidth}  }
\caption{Cluster of network of tRNA genes of different anticodons. They are
segments of the topology of 3420 tRNA genes. (a) Composing 96 vertices and
97 edges, similarity degree $S_0$ is 60, contain anticodons: ACG (solid
circle) and CCA (hollow circle); (b) Composing 226 vertices and 227 edges,
similarity degree $S_0$ is 60, contain anticodons: TAG (solid circle) and
TGA (hollow circle). }
\label{figure5}
\end{figure}
}

\end{document}